\documentclass[a4paper]{article}

\usepackage{INTERSPEECH2021}
\usepackage{tikz}
\usetikzlibrary{shapes.geometric, arrows}
\usetikzlibrary{fit, backgrounds, positioning, shadows}
\title{Scaling Laws for Acoustic Models}
\name{Jasha Droppo, Oguz Elibol}
\address{Amazon Alexa}
\email{\{drojasha, oelibol\}@amazon.com}

\begin{document}
\ninept

\tikzstyle{process} = [rectangle, minimum width=3cm, minimum height=0.75cm, text centered, draw=black]
\tikzstyle{arrow} = [thick, ->, >=stealth]
\tikzstyle{background}=[rectangle,
	fill=gray!10,
	inner sep=0.2cm,
	anchor=north,
	rounded corners=5mm]
	\tikzset{%
	cascaded/.style = {%
	  general shadow = {shadow scale = 1, shadow xshift = -1ex, shadow yshift = 1ex, draw,	thick, fill = white},
	  general shadow = {shadow scale = 1, shadow xshift = -.5ex, shadow yshift = .5ex, draw, thick,fill = white},
	  fill = white, draw, thick,
	  minimum width = 1.5cm,
	  minimum height = 2cm}}

\maketitle
\begin{abstract}
There is a recent trend in machine learning to increase model quality by growing
models to sizes previously thought to be unreasonable. Recent work has shown
that autoregressive generative models with cross-entropy objective functions exhibit smooth power-law
relationships, or scaling laws, that predict model quality from model size,
training set size, and the available compute budget. These scaling laws allow
one to choose nearly optimal hyper-parameters given constraints on available
training data, model parameter count, or training computation budget. In this
paper, we demonstrate that acoustic models trained with an auto-predictive
coding loss behave as if they are subject to similar scaling laws. We extend previous work to
jointly predict loss due to model size, to training set size, and to the
inherent ``irreducible loss'' of the task. We find that the scaling laws
accurately match model performance over two orders of magnitude in both model size and training set size, and make
predictions about the limits of model performance.

\end{abstract}

\section{Introduction}

In the field of language modeling, it has been shown that generative models have
predictable and favorable scaling properties with respect to training set size,
model size, and available compute budget\cite{2020arXiv200108361K,
2020arXiv201014701H}. It has also been shown that, as the models become larger
and more accurate, less task-specific fine-tuning is necessary to achieve
state-of-the-art results\cite{brown2020language}.

Consequently, the field of natural language processing is in the middle of an
``arms race'' to build the largest and most powerful models. It started with
smaller models like GPT\cite{GPT} and Bert\cite{Bert}, which had hundreds of
millions of parameters. Today, state-of-the-art models like
SWITCH-C\cite{SWITCHC} have more than 1.5 trillion parameters.

Meanwhile, in the field of acoustic modeling of speech, there has been growing
interest in improving automatic speech recognition by using bulk untranscribed
audio for unsupervised pre-training. Successful techniques use either
autoregressive predictive coding (APC) \cite{2020-DeCoAR2, 2020-APC, 2020-VQAPC}
or contrastive predictive coding (CPC) \cite{2020-wav2vec2, 2018-CPC,
2020arXiv201011430X} to pre-train an acoustic model from a large corpus of
unlabeled speech. Once the pre-training is complete, the models are used to
generate features for downstream modeling tasks, such as automatic speech
recognition.

\interfootnotelinepenalty=1000000 If acoustic models trained with predictive
coding exhibit similar scaling behavior as the autoregressive generative models
studied in~\cite{2020arXiv201014701H}, then we can use this analysis to discover
best practices for scaling and training acoustic models that are much larger and
more accurate than today's state-of-the-art. By accurately predicting the amount
of data, number of parameters, or length of compute, we eliminate these factors
from potentially expensive hyperparameter tuning on the models we build. By
understanding whether our model's performance (development-set loss) is being
limited by a lack of data or by the model size, we can make intelligent
decisions about how to improve them.

This paper tests this premise and develops a generalization of previous work
that accounts for the interplay between limited training data, limited model
parameters, and the development-set loss of a converged model.

We find that these acoustic models do exhibit smooth power-law relationships
with respect to the training constraints, that training to convergence is
inefficient, that data requirements scale sub-linearly with model size, and that
larger models are more efficient learners as was shown for language models
in~\cite{2020arXiv200108361K}.

This paper is organized as follows. In Section~\ref{sec:models}, we describe the
acoustic models that we test, our training procedure, and our parameter scaling
strategy. In Section~\ref{sec:converged}, we review previously published results
on model scaling, and fit them to measurements of fully trained acoustic models.
Section~\ref{sec:efficient} highlights the advantage of not training to
convergence, and or conclusions are presented in Section~\ref{sec:conclusion}.

\section{Experimental Setup}
\label{sec:models}

\begin{figure}[htp] 
	\begin{center}
	\begin{tikzpicture}[node distance=1.25cm]
		\node (speech) [text width=1cm] {Input Speech};
		\node (encoder) [process, right of=speech, anchor=north, rotate=90] {Encoder};
		\node (context) [process, right of=encoder, anchor=north, rotate=90] {Context};
		\node (prediction) [process, right of=context, anchor=north, rotate=90] {Prediction};
		\node (loss) [right of=prediction, text width=2cm, xshift=1cm] {Predicted Speech};
		\draw [arrow] (speech) -- (encoder);
		\draw [arrow] (encoder) -- (context);
		\draw [arrow] (context) -- (prediction);
		\draw [arrow] (prediction) -- (loss);
	\end{tikzpicture}
	\vspace{-0.75cm}
\end{center}
	\caption{APC model structure.}
	\label{fig:model}
\end{figure}
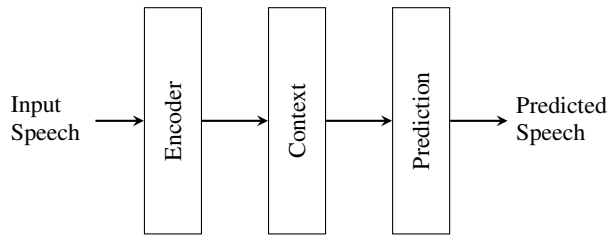

The two models evaluated in this work share the same basic auto-predictive
coding (APC) design, shown in Figure~\ref{fig:model}. Both models share the same
encoder and prediction structures, with different context modules.

In this work, the encoder module is a dense affine embedding layer that projects
each input feature vector into the layer width of the context module, without
additional frame-stacking or decimation.

The context module is a sequence-to-sequence model that converts the encoded
input sequence into a sequence of context vectors. We experiment with two
different designs for the context model: the LSTM and the Transformer. To
maintain causality, the LSTM are uni-directional and the Transformer uses
masking to prevent the network from using future frames.

The prediction module uses the output of the context module to predict and
reconstruct ten consecutive input feature vectors. It does this with ten
separate prediction networks. Each one consists of a linear projection to 512
units, a ReLU nonlinearity, and a linear projection to the input data dimension.
The model is trained to minimize the sum of the L1 loss between each of the ten
prediction networks and the ten contiguous input feature vectors. Following the
design in~\cite{DeCoAR}, the first of these feature vector targets is identical
to the context vector's latest input.

The difference between this APC design and a more traditional autoencoder is
that in an autoencoder, the network distills and recreates the input, and in an
APC design, the network predicts the future from the past.

The LSTM context module is a stack of uni-directional LSTM layers. Each of the
$L$ layers incorporates layer normalization, and every layer but the first
employs a residual structure. The complexity of this context module, both in
terms of number of parameters $N$ and the number of multiplications per frame of
input data performed during inference $M$ are shown in
Table~\ref{table:lstm_properties}.
The values are a function of $u$, the number of units in the LSTM.

\begin{table}[htp]
	\begin{center}
		\caption{LSTM layer parameter count $N$ and inference multiplication count $M$.}
	\label{table:lstm_properties}
	\begin{tabular}{lrr}
		Computation & $N$ & $M$
		\\ \hline
		Input, Output, Forget Gates & $3u(2u+1)$ & $3u(2u+1)$
		\\
		Cell Update & $u(2u+1)$ & $u(2u+1)$
		\\
		Layer Norm & $2u$ & $u$
		\\ \hline
		Total & $4u(2u+1)$ & $u(8u+5)$
	\end{tabular}
\end{center}
\end{table}

The Transformer context model is a stack of Transformer encoder layers. Each
layer comprises layer norm, masked multi-headed self-attention, a second layer
norm, and a then a point-wise feed-forward network. Both the self-attention and
the feed-forward network employ a residual connection. The number of parameters
$N$ and the number of multiplications per input frame during inference $M$ are
given in Table~\ref{table:xfmr_properties}. Unlike the LSTM, the computational
complexity of the transformer-based context module is dependent on both the
number of units in the model $u$ and the attention component's context length
$n_\mathrm{c}$.

\begin{table}[htp]
	\begin{center}
		\caption{Transformer encoder layer parameter count $N$ and forward-pass multiplication count $M$.}
	\label{table:xfmr_properties}
	\begin{tabular}{lrr}
		Computation & $N$ & $M$
		\\ \hline

		Two Layer Norms & $4u$ & $2u$ \\
		Att. Embeddding & $3u(u+1)$ & $3u(u+1)$ \\
		Self Attention & $0$ & $2n_\mathrm{ctx}u$ \\
		Att. Projection & $u(u+1)$ & $u(u+1)$ \\
		Feed-Forward & $u(8u+5)$ & $u(8u+5)$ \\
		\hline
		Total & $u(12u+13)$ & $u(12u+2n_\mathrm{ctx}+11)$
	\end{tabular}
\end{center}
\end{table}

The LSTM exhibits $O(u^2)$ complexity, regardless of the input data. The
Transformer exhibits $O(u^2)$ complexity for short context lengths, and $O(u)$
for long context lengths. Specifically, when $n_\mathrm{c} < 6u$, the term
$12u^2$ dominates the Transformer multiplication count, and when
$n_\mathrm{c}>6u$, the term $2un_\mathrm{c}$ becomes more important.


\subsection{Training}

To train the models, we used an Adam optimizer with a warm hold decay learning
rate scheduler. The learning rate was ramped up from $1\times 10^{-4}$ to
$2\times 10^{-4}$ over 3,000 steps, held there until 50,000 steps were complete,
and then exponentially decayed to hit $1\times 10^{-5}$ at 150,000 steps. Each
step consisted of a 64-example mini-batch of speech sequences, with a feature
dimension of 64 and roughly bucketed sequence lengths. Loss function values were
computed on a held-out development set every 25,000 steps during training.

All acoustic data used in this paper was drawn from a 23 thousand hour
corpus of untranscribed, de-identified, far-field, English voice command and voice query speech
collected from home environments. This data is presented to the network as a series of log-Mel frequency
filterbank feature vectors, at a rate of 100 vectors per second of audio. Although this data is not publicly available,
the authors believe that the phenomena described in this paper should apply to
any similar set of speech recordings.

\subsection{Model Scaling Strategy}

Although one could scale model size by independently choosing the number of
layers $n_\mathrm{layer}$ and number of units $u$ in the context module, we
chose to scale the models by coupling the two hyper-parameters with a fixed
aspect ratio.

Preliminary experimentation indicated that for a given number of parameters, the
LSTM model performs best with a constant aspect ratio of $u = 256 n_{layer}$.
Similarly, the transformer model performs best with an aspect ratio of $u=64
n_{layer}$. These ratios are used exclusively throughout this paper.

\section{Scaling of Converged Models}
\label{sec:converged}

This section describes the relationship between the number of model parameters
$N$, the amount of training data $D$, and the model loss of a fully trained
model $L(N,D)$.

\subsection{Limited Data or Limited Model Parameters}
\label{sec:or}

It was empirically demonstrated in \cite{2020arXiv200108361K} and
\cite{2020arXiv201014701H} that, for several generative modeling tasks, the loss
of autoregressive Transformer models behave predictably when their performance
is limited by either the number of model parameters $N$ or the amount of
training data $D$. Specifically, loss is predictable according to a power-law
plus constant formulation as in Eq.~\ref{eq:nd_separate}.
\begin{equation}
	\begin{matrix}
	L(D) = L_{\infty} + \left(\frac{D_C}{D}\right)^{\alpha_D},
	&
	L(N) = L_{\infty} + \left(\frac{N_C}{N}\right)^{\alpha_N}
	\end{matrix}
	\label{eq:nd_separate}
\end{equation}

The parameter $D_C$ represents the critical value of $D$, where the contribution
of limited data to the loss function is equal to 1.0. The parameter $\alpha_D$
controls the relationship between increased data and reduced loss. The
parameters $N_C$ and $\alpha_N$ perform similar functions in relating $N$ to
$L(N)$. The constant $L_{\infty}$ is the ``irreducible loss'' that is
independent of either $D$ or $N$.

\begin{figure}[htp]
    \begin{center}
        \input{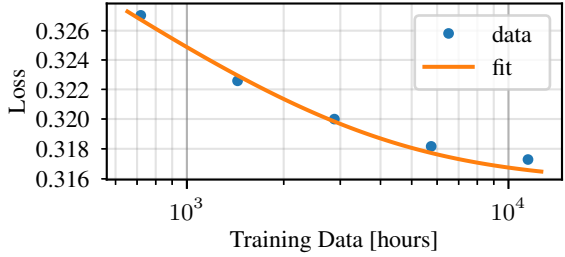}
		\vspace{-1cm}
    \end{center}
    \caption{When model performance is limited only by available training data
		the relationship between training data and loss is linear on a log-log
		plot.}
	\label{fig:limited_data}
\end{figure}

To demonstrate the effect of limited $D$ on loss, we trained several models that
were likely to be performance-limited by $D$. Five instances of 11-layer
Transformer models were on differently sized subsets of the training data. The
training data sizes $D$ were chosen to be exponentially spaced from 180 hours to
11,500 hours. The loss values for these models are marked with dots in
Figure~\ref{fig:limited_data}. The parametric $L(D)$ from
Eq.~\ref{eq:nd_separate} is shown with constants $D_C=7.350\times 10^{-23}$ hours,
$\alpha_D=0.01946$, and $L_\infty=0.316$.

The small value for $D_C$ indicates that almost no data is necessary
to bring the objective function below 1.0. This makes sense for our data, where
if the model did nothing but predict the mean spectral value, the objective
function would already be less than 1.0.

\begin{figure}[htp]
    \begin{center}
        \input{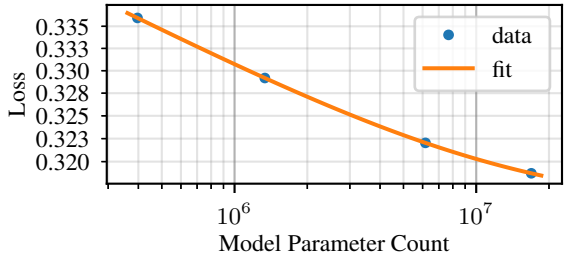}
		\vspace{-1cm}
    \end{center}
    \caption{When model performance is limited by model size, the relationship
		between parameter count and loss is linear on a log-log plot.}
	\label{fig:limited_parameters}
\end{figure}

To demonstrate the effect of limited $N$ on loss, we trained several models that
were likely to be performance-limited by $N$.
Five models were trained on 23
thousand hours of data, with model size restricted to 2, 3, 5, 7, and 11 layers.
The resulting loss values are marked with dots in
Figure~\ref{fig:limited_parameters}, together with the parametric $L(N)$ with
$N_C=9.410\times 10^{-25}$ parameters, $\alpha_N=0.01601$, and $L_\infty=0.316$. Note that
the value used for irreducible loss here in $L(N)$ is identical to the one used
for $L(D)$.
The small value for $N_C$ indicates that loss values less than 1.0 are
achievable with zero parameters in the context network.

Together, these relationships describe how to improve model performance when it
is being limited either by $D$ or $N$.

At less than about one thousand hours of training data, the 11-layer transformer
is data-limited. With the estimated value of $\alpha_D$, a five percent reduction
in loss requires a 14.0-fold increase in training data $D$.

At less than about two million parameters, the models trained on a large amount
of data are parameter-limited. With the estimated value of $\alpha_N$, a five
percent reduction in loss requires a 24.6-fold increase in model size $N$.

The ratio of $\alpha_N$ to $\alpha_D$ describes how quickly $D$ should increase
to support an increase in $N$. If $D$ scales more slowly, then the model
performance will eventually be limited by $D$ and the increase in $N$ will be
wasted. For our data, $\frac{\alpha_N}{\alpha_D}<1$, which means data
requirements scale sub-linearly with model parameter count. Specifically, with
every doubling of $N$, $D$ should increase by a factor of 1.77.

If our only constraint is model size $N$, we should choose $D$ large enough so
that its contribution to the loss is at least one order of magnitude smaller
than $L(N)$. For our data, this implies a lower limit of $D > 0.0436 N ^ {0.8230}$.

\subsection{Limited Data and Limited Model Size}

In~\cite{2020arXiv200108361K}, the authors demonstrated that when limited by
both data and model size, but with unlimited computational power, the two
previous relationships can be combined into a single scaling law,
Eq.~\ref{eq:ND_simple}. Unfortunately, this equation was developed for language
models where $L_\infty$ can be ignored. The authors of
\cite{2020arXiv201014701H} did not present an analogous relationship that
incorporates the concept of irreducible loss.
\begin{equation}
	L(N,D) = \left[\left(\frac{N_C}{N}\right)^{\frac{\alpha_N}{\alpha_D}}+\left(\frac{D_C}{D}\right)\right]^{\alpha_D}
\label{eq:ND_simple}
\end{equation}

To adapt this relationship to account for irreducible loss, we introduce the
following generalization, which is equivalent to Eq.~\ref{eq:ND_simple} with the
addition of an ``irreducible loss'' term and a slight change of variables.
\begin{equation}
	L(N,D) = \left[\left(L_{\infty}\right)^\frac{1}{\alpha}+\left(\frac{N_C}{N}\right)^\frac{\alpha_N}{\alpha}+\left(\frac{D_C}{D}\right)^\frac{\alpha_D}{\alpha}\right]^{\alpha}
	\label{eq:lnd}
\end{equation}

To test this scaling law's ability to describe model performance, we trained
twenty-one Transformer based models. We used a variety of $n_\mathrm{layer}$
from 2 to 11, a fixed aspect ratio of $u=64n_\mathrm{layer}$, and training data
set sizes from 134 hours to 23 thousand hours of speech. Each model was trained
until the development set loss began to increase, or until 1.5 million parameter
updates were complete.

\begin{figure}[htp]
    \begin{center}
        \input{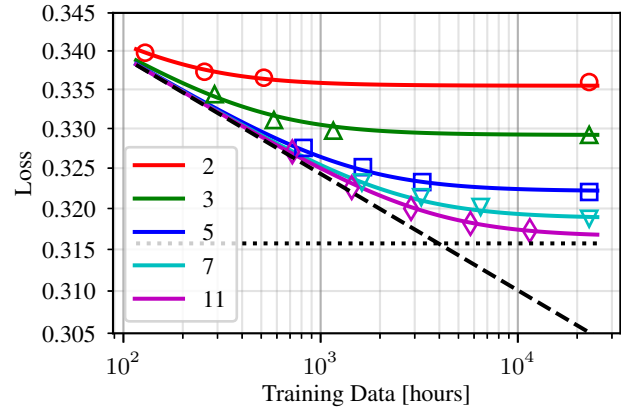}
		\vspace{-1cm}
    \end{center}
    \caption{The converged loss of several Transformer models with 2 through 11 layers,
	plotted as a
    function of training data $D$. Each marker represents a different experiment
    across $n_\mathrm{layer}$ and $D$. Lines are drawn for each model size
    representing the prediction of Eq.~\ref{eq:lnd}. The diagonal dashed line
    represents a performance frontier imposed by limited data, and the
    horizontal dotted line represents the irreducible loss.}
	\label{fig:combined_scaling}
\end{figure}

The loss values for these models are represented by markers in
Figure~\ref{fig:combined_scaling}. The performance barrier due to limited
training set size is shown as a diagonal dashed line. Regardless of model size,
we do not expect any model will be able to surpass this limit. The performance
barrier due to irreducible loss is represented as a horizontal dotted line. If
Eq.~\ref{eq:lnd} is correct, no model can be better than this limit, regardless of
model size or training data set size.

The smooth curves in Figure~\ref{fig:combined_scaling} are computed using
Eq.~\ref{eq:lnd} with the same constants reported above with the addition of
$\alpha=0.01363$. It is clear that the model of Eq.~\ref{eq:lnd} is a good match
to our data.

\section{Efficient Scaling Properties}
\label{sec:efficient}

Both \cite{2020arXiv200108361K} and \cite{2020arXiv201014701H} showed that for a
broad class of models, training to convergence is a computationally inefficient
way to maximize model performance. Instead, they demonstrated that for a given
loss, the most efficient way to train a model to that loss is to train a larger
model using fewer model update steps. We repeat this analysis for acoustic
models.

\begin{figure}[htb]
    \begin{center}
        \input{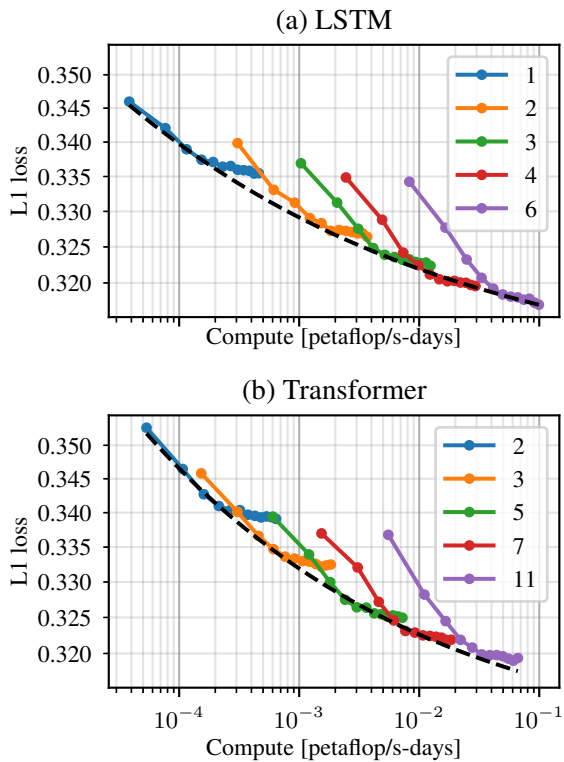}
		\vspace{-1cm}
    \end{center}
    \caption{Development set loss for both LSTM and Transformer models for models with the indicated number of layers.
	The dashed line represents the computationally efficient frontier defined in Eq.~\ref{eq:compefficient}.
	}
	\label{fig:power_law}
\end{figure}

Figure~\ref{fig:power_law} shows the evolution of development-set loss during
training for both LSTM and Transformer based models. Each curve represents an
experiment with a different $N$. Each experiment is stopped after 300,000 model
updates, which represents less than one full pass over the training data. For
each point, the x-coordinate is the approximate number of multiplications and
additions executed during training, following the conventions established
in~\cite{2020arXiv200108361K}.


Regardless of model size, no experiment surpasses the power-law plus constant
fit shown as a dotted line. This line is the compute-efficient frontier, and has
a similar form to Eq.~\ref{eq:nd_separate}.
\begin{equation}
	L(C) = L_\infty + \left(\frac{C_C}{C}\right)^{\alpha_C}
	\label{eq:compefficient}
\end{equation}

When a model reaches $L(C)$, it means that a different model with enough
capacity, but with fewer parameters, would need more computation and more data
to reach the same loss value. Alternatively, a model with more parameters would
need more computation and less data to reach the same loss value.

Where curves for two experiments meet, it is an indication that the same amount of compute
can reach the given loss value through two different methods. One can either use
more parameters and fewer data, or use fewer parameters and more data.

The constant $L_\infty$ is 0.306 in both figures. This represents a shared
asymptote between the LSTM and Transformer systems, which will never be
surpassed, regardless of the computational or data budget. The fact that the
same asymptote applies to both systems hints that irreducible loss is indeed a
fundamental property of the data and not the model. Additionally, this constant
is similar to the value found in Section~\ref{sec:or}. The authors suspect that
the constants should be identical, but our precision in measuring it is limited.

The LSTM models exhibit a compute-efficient frontier with a slope of -0.167. A
doubling of computation yeilds a $10.9\%$ reduction in objective function. A
halving of objective function would come with a $63.5$ fold increase in
computation.

The slope of the compute-efficient frontier for Transformer models is -0.197.
When computation is increased by a factor of r, then the reducible loss will be
changed by a factor of $r^{-0.197}$. At that rate, a doubling of computation
yields a $12.7\%$ reduction in objective function. A halving of objective
function would come with a $33.7$ fold increase in computation.

The difference in slope between the LSTM and Transformer experiments indicate
that the Transformer architecture makes more efficient use of increased model
parameters and increased training data. Although LSTM is superior to transformer
at smaller model sizes, as the model size grows, and these trends continue, the
transformer will eventually be more efficient.


Finally, the experimental data show that larger models learn more quickly from
the same amount of data. Each of the points plotted in
Figure~\ref{fig:power_law} represent the consumption of an additional 25
thousand minibatches of training data. At the first point, second, or third,
each model has processed the same data, but the larger models have achieved
better accuracy on the held-out development set.

\section{Conclusion}
\label{sec:conclusion}


In this paper, we demonstrated that the effect of limited model parameters and
limited training data on the quality of APC acoustic models follows a power-law
given by Eq.~\ref{eq:lnd}. This extends previous results in language modeling
and recurrent generative modeling.
The relationship can be used to predict the amount of data needed to support a
given model size, or to predict the proper model size for a given amount of
data.

For our data, we have shown that the irreducible loss $L_\infty$ is the same,
whether the model uses Transformer or LSTM components. Furthermore, under two
different power-law relationships, the $L_\infty$ is similar. Together, this
suggests that it is a fundamental property of the task and data, but not the
model. Although speculative, authors believe that the irreducible loss derived
from such scaling studies can translate into a measure of information content,
or quality of the training data. One study that can be performed, beyond the
scope of this work, is to add controlled amount of noise and study the effect on
the irreducible loss. Future work should tie $L_\infty$ to the intrinsic
difficulty in modeling a set of acoustic data, similar to perplexity
measurements in language modeling tasks.

This paper has implicitly assumed that improvements in L1 loss are desirable.
Future work will relate L1 loss to performance on a downstream acoustic modeling
task.


\bibliographystyle{IEEEtran}

\bibliography{mybib}

\end{document}